\documentclass[aps,prd,superscriptaddress,10pt]{revtex4-2}
\usepackage{color}
\usepackage{xcolor}
\usepackage{graphicx}
\usepackage[utf8]{inputenc} 
\usepackage[normalem]{ulem}
\usepackage{amsmath}
\usepackage{amssymb}
\usepackage{float}
\usepackage{comment}
\usepackage{slashed}

\definecolor{darkred}{rgb}{0.6,0,0}
\usepackage[colorlinks=true,linkcolor=darkred,citecolor=darkred,urlcolor=darkred, pdfborder={0 0 0}]{hyperref}
\usepackage{silence}
\usepackage{tikz}
\usepackage[compat=1.1.0]{tikz-feynman}
\tikzfeynmanset{warn luatex=false}

\usepackage{orcidlink}

\begin{document}


\title{\boldmath \color{darkred} Dirac neutrinos and gauged lepton number}

\author{A. E. C\'{a}rcamo Hern\'{a}ndez~\orcidlink{0000-0002-2421-2732}}
\email{antonio.carcamo@usm.cl}
\affiliation{Universidad T\'{e}cnica Federico Santa Mar\'{\i}a, Casilla 110-V, Valpara\'{\i}so, Chile}
\affiliation{{Centro Cient\'{\i}fico-Tecnol\'ogico de Valpara\'{\i}so, Casilla 110-V,
Valpara\'{\i}so, Chile}}
\affiliation{Millennium Institute for Subatomic Physics at High-Energy Frontier (SAPHIR),
Fern\'andez Concha 700, Santiago, Chile}

\author{Andr\'{e}s Enr\'{i}quez~\orcidlink{0009-0002-3176-0208}}
\email{jjandres1997@gmail.com}
\affiliation{Departamento de F\'isica, Divisi\'on de Ciencias e Ingenier\'ias, Campus Le\'on, Universidad de
  Guanajuato, Loma del Bosque 103, Lomas del Campestre, Le\'on 37150, Guanajuato, Mexico}

\author{Sergey Kovalenko~\orcidlink{0000-0002-8518-2282}}
\email{sergey.kovalenko@unab.cl}
\affiliation{Departamento de Astronom\'ia y F\'isica, Universidad Andres Bello, \\
   Sazi\'e 2212, Piso 7, Santiago, Chile}
\affiliation{Millennium Institute for Subatomic Physics at High-Energy Frontier (SAPHIR),
Fern\'andez Concha 700, Santiago, Chile}
\author{Eduardo Peinado~\orcidlink{0000-0003-3803-3353}}\email{epeinado@fisica.unam.mx}
\affiliation{Instituto de Física, Universidad Nacional Autónoma de México
Ciudad de México, C.P. 04510, Mexico}
\author{Carlos Alberto Vaquera-Araujo~\orcidlink{0000-0001-8578-9263}}\email{vaquera@fisica.ugto.mx}
\affiliation{Secretar\'ia de Ciencia, de Humanidades, Tecnolog\'ia e Innovaci\'on, Insurgentes Sur 1582. Colonia Cr\'edito Constructor, Benito Ju\'arez 03940, Ciudad de M\'exico, Mexico}
\affiliation{Departamento de F\'isica, Divisi\'on de Ciencias e Ingenier\'ias, Campus Le\'on, Universidad de
  Guanajuato, Loma del Bosque 103, Lomas del Campestre, Le\'on 37150, Guanajuato, Mexico}
\affiliation{Dual $CP$ Institute of High Energy Physics, Colima 28045, Colima, Mexico}

\begin{abstract}
We propose the first scotogenic neutrino mass model with gauged lepton number $U(1)_L$, which is spontaneously broken by three units $\Delta L=3$ down to  
a residual discrete gauge symmetry $\mathbb{Z}_6$. The latter guarantees that neutrinos acquire tiny Dirac masses via a one-loop scotogenic mechanism, simultaneously stabilizing the lightest electrically neutral  particle with nontrivial charge under the preserved $\mathbb{Z}_6$ symmetry.  In our model there is a scalar particle identified as weakly interacting massive particle dark matter (DM) candidate. We analyzed its compatibility with the existing data on direct  DM  detection experiments and the DM relic abundance. We also address charged lepton flavor violating decays in our model and find that their predicted rates are within the reach of current experimental sensitivity.
\end{abstract}


\maketitle
\noindent

\section{Introduction}
\label{Sect:intro}

Although the Standard Model (SM) has been remarkably successful in describing strong and electroweak (EW) interactions, it leaves several fundamental questions unresolved, including the origin and smallness of neutrino masses. Furthermore, the SM does not contain a viable, stable candidate for dark matter (DM). Resolving these shortcomings necessitates the introduction of physics beyond the SM. A well-motivated framework in this context is the scotogenic model \cite{Tao:1996vb,Ma:2006km}, in which the generation of neutrino masses is intrinsically linked to a DM particle. Within this scenario, active neutrino masses arise radiatively at the one-loop level, while the lightest electrically neutral particle participating in the loop serves as a weakly interacting massive particle (WIMP). Its stability is ensured by an \emph{ad hoc} global $\mathbb{Z}_2$ symmetry, thereby providing a coherent explanation for both neutrino mass generation and the DM puzzle.

An appealing alternative to the original scotogenic paradigm is that the symmetry responsible for stabilizing DM is not imposed by hand, but instead arises naturally from a symmetry intimately connected to neutrinos.
It is well known that in the minimal formulation SM, the conservation of both baryon number $B$ and lepton number $L$  is an accidental feature of the theory. This means that $B$ and $L$ are symmetries in the SM because of the specific interactions and particle content of the theory, but there is no underlying dynamical mechanism that enforces this conservation. 
One of the simplest and most widely studied extensions of the SM involves the promotion of $\mathrm{U}(1)_{B-L}$ to a gauge symmetry. The most straightforward approach is to include three right-handed neutrinos, which are singlets under the Standard Model gauge group and carry a lepton number of $+1$. This minimal choice ensures that the theory remains anomaly-free. A lesser-known fact is that the individual gauging of $B$ and $L$ is also possible and has been studied in \cite{FileviezPerez:2010gw, Duerr:2013dza,FileviezPerez:2024fzc}, where a set of new fermions is included to cancel the gauge anomalies. 
In this paper, we focus on an SM extension with gauged lepton number $L$ and global accidental baryon number $B$. The gauging of lepton number introduces several interesting and novel features to the theory, like the existence of a neutral gauge boson that couples to leptons but not to quarks; new mechanisms for neutrino masses and mixing \cite{Debnath:2023akj,Debnath:2024vpf,Debnath:2025rbu}; new sources of lepton flavor violation, leading to potentially observable processes such as rare decays \cite{Debnath:2025dvl}; and distinctive collider signatures \cite{Chang:2018nid,Kara:2025dxl}. The particle interactions in this class of models can be constrained by future neutrino-electron scattering experiments \cite{Chakraborty:2021apc}.

In this work, we propose the first scotogenic model for a theory with gauged lepton number. In this model, $L$ is spontaneously broken by three units, and light Dirac neutrino masses are generated at one-loop. Due to the introduction of mediators with fractional lepton number in the scotogenic loop, a remnant discrete gauge symmetry $\mathbb{Z}_{6}$ stabilizes the lightest neutral particle with non trivial transformation properties under this symmetry, which can be identified as a WIMP DM candidate. Furthermore, that preserved $\mathbb{Z}_{6}$ discrete symmetry guarantees the radiative nature of the one-loop level scotogenic mechanism that produces the tiny active neutrino masses. In contrast, the model in Ref. \cite{Kara:2025dxl}, where a leptophilic symmetry is gauged, generates neutrino masses from a tree-level type I seesaw mechanism.

The paper is organized as follows: In Sec. \ref{Sect:model} we present the matter content of the model and its transformation
properties under the gauge and residual symmetries. Sec. \ref{Sect:boson} is devoted to the identification of the physical scalars and gauge
bosons of the model. The scotogenic mechanism for light Dirac active neutrino masses is presented in Sec. \ref{Sect:scotogenic}. We study the viability of a WIMP DM candidate
identified with the lightest neutral scalar in the dark sector in Sec. \ref{Sect:Dark}, and the
phenomenology of charged lepton flavor violation is analyzed in Section \ref{Sect:cLFV}. Finally we summarize our results in Sec.
\ref{Sect:Conclusions}.

\section{The Model}
\label{Sect:model}

We consider an extension of the inert doublet model where the scalar sector is enlarged by the inclusion of two gauge singlet scalars, and the SM fermion spectrum is augmented by three right-handed neutrinos and a family of vectorlike leptons under the SM gauge symmetry. In our model, the tiny active neutrino masses are generated from a one-loop level radiative seesaw mechanism mediated by three Dirac neutral leptons with fractionally $L$ charges and the inert scalar fields. The model is based on the gauge symmetry
\begin{eqnarray}
\label{eq:GaugeGroup-1}
\mathcal{G}\equiv \mathrm{SU}(3)_{C}\otimes \mathrm{SU}(2)_{%
W}\otimes \mathrm{U}(1)_{Y}\otimes \mathrm{U}(1)_{L},
\end{eqnarray}
and undergoes the following spontaneous symmetry breaking (SSB) pattern
\begin{eqnarray}
\mathcal{G}\quad&&{\xrightarrow{w}}\quad\mathrm{SU}(3)_{C}\otimes \mathrm{SU}(2)_{W}\otimes \mathrm{U}(1)_{Y}\otimes \left(\mathbb{Z}_{6}\subset U(1)_L\right)\quad
{\xrightarrow{v}}\quad\mathrm{SU}(3)_{C}\otimes \mathrm{U}(1)_{Q}\otimes \mathbb{Z}_{6} ,\label{SB}
\end{eqnarray}%
with $w$ and $v=246$ GeV as the scales of the lepton number symmetry and the electroweak symmetry breaking scales, respectively. Throughout this work, we will assume the hierarchy $w>v$.

The field content of the model and their transformation properties under $\mathcal{G}$, as well as their transformations under the residual $\mathbb{Z}_{6}$ discrete symmetry, are displayed in Table \ref{tab3211}. The inclusion of the new leptonic fields is crucial for ensuring the cancellation of the chiral anomalies, 
as shown in Appendix \ref{App:anomaly}. The extra fermion fields comprise a vectorlike family of leptons with arbitrary lepton numbers $\ell-3$ and $\ell$, two right-handed neutrinos with lepton number $4$ and one with $-5$ \cite{Montero:2007cd}, and vectorlike fields $S_i$ ($i=1,2,3$) with fractional lepton number $1/2$ that serve as mediators in the scotogenic mechanism for neutrino mass generation at one loop. Without the inclusion of the vectorlike neutral leptons $S_i$, the scotogenic mechanism cannot be implemented, yielding massless active neutrinos. The scalar sector is extended with respect to the SM with the inclusion of an inert doublet $\eta$ whose lepton number is $-1/2$ and an electroweak singlet $\sigma$ with lepton number $-7/2$ to close the scotogenic neutrino mass loop, as well as a singlet field $\phi$ responsible for triggering the spontaneous breakdown of the lepton number symmetry. 
\begin{table}[ht]
\centering
\begin{tabular}{|c|c|c|c|c|c|c|c|}
\hline
\hspace{0.2cm} Field \hspace{0.2cm} & \hspace{0.2cm}SU(3)$_C$ \hspace{0.2cm}
& \hspace{0.2cm} SU(2)$_W$ \hspace{0.2cm} & \hspace{0.2cm}U(1)$_Y$ \hspace{%
0.2cm} & \hspace{0.2cm}U(1)$_{L}$ \hspace{0.2cm} &\hspace{0.2cm} $\mathbb{Z}_6$ 
\hspace{0.2cm}   \\ 
\hline\hline
$Q_{iL}$ & \textbf{3} & $\textbf{2}  $ & $\frac{1}{6}$ & $0$ & $1$ \\ 
$u_{iR}$ & \textbf{3} & \textbf{1} & $\frac{2}{3}$ & $0$ & $1$\\ 
$d_{iR}$ & \textbf{3} & \textbf{1} & $-\frac{1}{3}$ & $0$ & $1$\\
$L_{iL}$ & \textbf{1} & \textbf{2} & $-\frac{1}{2}$ & $1$ & $\omega^2$\\ 
$e_{iR}$ & \textbf{1} & \textbf{1} & $-1$ & $1$  & $\omega^2$\\ 
\hline
$\nu_{aR}$ & \textbf{1} & \textbf{1} & $0$ & $4$  & $\omega^2$ \\
$\nu_{3R}$ & \textbf{1} & \textbf{1} & $0$ & $-5$  & $\omega^2$ \\
$L'_{L}$ & \textbf{1} & \textbf{2} & $-\frac{1}{2}$ & $\ell-3$ & $\omega^{2\ell}$\\ 
$e'_{R}$ & \textbf{1} & \textbf{1} & $-1$ & $\ell-3$  & $\omega^{2\ell}$ \\ 
$n_{R}$ & \textbf{1} & \textbf{1} & $0$ & $\ell-3$ & $\omega^{2\ell}$\\
$L''_{R}$ & \textbf{1} & \textbf{2} & $-\frac{1}{2}$ & $\ell$ & $\omega^{2\ell}$\\ 
$e''_{L}$ & \textbf{1} & \textbf{1} & $-1$ & $\ell$  & $\omega^{2\ell}$ \\  
$n_{L}$ & \textbf{1} & \textbf{1} & $0$ & $\ell$ & $\omega^{2\ell}$\\ 
$S_{iL}$ & \textbf{1} & \textbf{1} & $0$ & $\frac{1}{2}$ & $\omega$\\ 
$S_{iR}$ & \textbf{1} & \textbf{1} & $0$ & $\frac{1}{2}$ & $\omega$\\ 
 \hline\hline
$H$ & \textbf{1} & \textbf{2} & $\frac{1}{2}$ & $0$ & $1$\\ 
\hline
$\phi$ & \textbf{1} & \textbf{1} & $0$ & $3$ & $1$\\ 
$\eta$ & \textbf{1} & \textbf{2} & $\frac{1}{2}$ & $-\frac{1}{2}$ & $\omega^5$ \\ 
$\sigma$ & \textbf{1} & \textbf{1} & $0$ & $-\frac{7}{2}$ & $\omega^5$ \\
 \hline
\end{tabular}%
\caption{Fermionic and scalar content and their transformations under the $\mathrm{SU}(3)_{C}\otimes \mathrm{SU}(2)_{%
W}\otimes \mathrm{U}(1)_{Y}\otimes \mathrm{U}(1)_{L}$ gauge and $\mathbb{Z}_{6}$ discrete symmetries. Here $a=1,2 $, $i=1,2,3$ and $\omega = e^{2\pi i/6}$. Notice that $Z_6$ is the remnant symmetry arising from the spontaneous breaking of the $\mathrm{U}(1)_L$ gauge symmetry.}
\label{tab3211} 
\end{table}

Following the spontaneous symmetry breaking (SSB) shown in Eq. (\ref{SB}), the $\mathrm{U}(1)_{\mathrm{L}}$ gauge symmetry is broken by three units of lepton number ($\Delta L=3$). Due to the fractional nature of the charges, this breaking results in the preservation of a residual $\mathbb{Z}_{6}$ discrete symmetry, which acts nontrivially on the new fermions $S_i$ and scalar mediators of the scotogenic neutrino mass mechanism. This ensures the absolute stability of the lightest electrically neutral particle within this sector, identifying it as a WIMP dark matter candidate. Furthermore, the preserved $\mathbb{Z}_{6}$ discrete symmetry guarantees the radiative nature of the one-loop level scotogenic mechanism that produces the tiny active Dirac neutrino masses.

\section{Boson Sector}
\label{Sect:boson}
\subsection{Scalars}
The scalar potential invariant under the gauge symmetries of the model is given by
\begin{eqnarray}\label{V}
    V &=& \sum_{s=H,\phi,\eta,\sigma } \left[ \mu_s^2 (s^\dagger s) + \lambda_s (s^\dagger s)^2\right]  + \lambda_{H \eta} (H^\dagger H)(\eta^\dagger \eta) +  \lambda^\prime_{H\eta}(H^\dagger \eta)(\eta^\dagger H) \\
    &&+ \lambda_{H \sigma}(H^\dagger H)(\sigma^* \sigma) + \lambda_{H \phi}(H^\dagger H)(\phi^* \phi)   + \lambda_{\eta \sigma} (\eta^\dagger \eta)(\sigma^* \sigma)  + \lambda_{\eta \phi} (\eta^\dagger \eta)(\phi^* \phi) \\
    &&+ \lambda_{\sigma \phi} (\sigma^* \sigma)(\phi^* \phi) +  \frac{\kappa}{\sqrt{2}}( \eta^\dagger H \sigma \phi  + \mathrm{H.c})   \,.  \nonumber
\end{eqnarray}
For the sake of simplicity, we consider a scenario of a $CP$ conserved scalar potential, which implies a real mass parameter $\kappa$. In that scenario of $CP$ conservation in the scalar potential, the $CP$ even states will not features mixings with the $CP$ odd states, then simplifying the analysis of the scalar mass spectrum. 

The scalar doublets can be expanded into components according to 
\begin{equation}
    H = \left( \begin{array}{c}
        G^+ \\
        \frac{v + h_1' +iG_1}{\sqrt{2}}
    \end{array} \right), \quad \eta = \left( \begin{array}{c}
        \eta^+ \\
        \eta^0
    \end{array} \right)= \left( \begin{array}{c}
        \eta^+ \\
        \frac{\eta^0_R +i\eta^0_I}{\sqrt{2}}
    \end{array} \right),
\end{equation}
and the field $\phi$ is written as 
\begin{equation}
    \phi = \frac{w + h_2' + i G_2}{\sqrt{2}}.
\end{equation} 
The minimization conditions of the scalar potential yields the following relations:
\begin{eqnarray}
    \mu_H^2&= -\frac{1}{2} \left(2v^2 \lambda_H + w^2 \lambda_{H \phi}\right), \\ 
    \mu_\phi^2 &= - \frac{1}{2} \left( v^2 \lambda_{H \phi} + 2 w^2 \lambda_\phi  \right). \nonumber
\end{eqnarray}

The charged scalars $G^\pm$ are the Goldstone bosons absorbed by the charged gauge fields $W^\mu_\pm$
and remain unmixed with the physical dark charged scalars $\eta^{\pm}$, with squared masses
\begin{equation}
    m^2_{\eta^\pm} = \mu_\eta^2+\frac{\lambda_{H \eta }
    v^2}{2}+\frac{\lambda_{\eta \phi}  w^2}{2}.
\end{equation}
After SSB, the electrically $CP$-odd scalars $G_1$ and $G_2$ remain massless and become the Goldstone bosons absorbed by linear
combinations of the neutral gauge fields $Z^\mu$ and $Z^{\prime\mu}$. In the $CP$-even scalar sector, a mixing between the electrically neutral fields $h_1'$ and $h_2'$ is induced after SSB and is given by the squared mass matrix 
\begin{equation}\label{higgses}
    M^2_{1} = \left(
        \begin{array}{cc}
         2 \lambda_H v^2 & \lambda_{H \phi} v w \\
         \lambda_{H \phi} v w & 2 \lambda_\phi  w^2 \\
        \end{array}
        \right),
\end{equation}
that can be diagonalized through the introduction of the physical states
\begin{equation}
    \left( \begin{array}{l}
        h_1 \\
        h_2
    \end{array} \right) = \left(\begin{array}{cc}
        \cos{\alpha} & \sin{\alpha} \\
        -\sin{\alpha} & \cos{\alpha}
    \end{array} \right) \left(\begin{array}{c}
        h_1' \\
        h_2'
    \end{array} \right)
\end{equation}
with mixing angle
\begin{equation}
    \tan{2\alpha} = \frac{v w \lambda_{H \phi}}{w^2 \lambda_\phi - v^2 \lambda_H},
\end{equation}
and masses
\begin{equation}
    m^2_{h_{1,2}} = \lambda_H v^2+\lambda_\phi 
    w^2\mp \sqrt{\lambda_{H \phi}^2 v^2 w^2+(\lambda_\phi w-\lambda_H v)^2}.
\end{equation}
The lightest neutral  $CP$-even field corresponds to the SM Higgs field, which in the limits $w\gg v$ or $\lambda_{H\phi}\rightarrow0$ recovers its familiar squared mass $m^2_{h_{1}} \approx 2\lambda_H v^2$.

After the electroweak and lepton number symmetry breaking, the mass term  $(\eta^0 , \sigma) M_\varphi^2 (\eta^0, \sigma)^\dagger$ leads to mixing of the  fields $\eta^0$ and $\sigma$, due to the nondiagonal entries of the squared mass matrix
\begin{equation}
    M^2_2 =  
   \frac{1}{2} \left(
        \begin{array}{cc}
        2\mu_\eta^2+\lambda_{H \eta}^{\prime} v^2+\lambda_{H\eta}
         v^2+\lambda_{\eta \phi } w^2 & \frac{\kappa v w}{ \sqrt{2}} \\
        \frac{\kappa v w}{ \sqrt{2}} & 2\mu_\sigma^2+\lambda_{H \sigma}  v^2+\lambda_{\sigma \phi}  w^2 \\
        \end{array}
    \right).
\end{equation}
Upon diagonalizing the squared mass matrix above, we can identify two complex electrically neutral scalars in the spectrum as
\begin{equation}
    \left(\begin{array}{c}
    \varphi_1^0 \\
    \varphi_2^0
    \end{array}\right)=\left(\begin{array}{cc}
    \cos \theta & \sin \theta \\
    -\sin \theta & \cos \theta
    \end{array}\right)\left(\begin{array}{c}
    \eta^0 \\
    \sigma
    \end{array}\right),
\end{equation}
with 
\begin{equation}\label{angle}
    \tan{2\theta} = \frac{\sqrt{2} \kappa v w}{2 (\mu_\eta^2-\mu_\sigma^2)+v^2
    (\lambda_{H\eta}^{\prime} +\lambda_{H\eta}-\lambda_{H\sigma})+w^2
    (\lambda_{\eta \phi} -\lambda_{\sigma \phi} )}.
\end{equation}
The mass eigenvalues of these states are given by 
\begin{eqnarray}\label{phys_mass}
    m^2_{\varphi^0_{1,2} } &=& \frac{1}{4} \Bigg\{2 (\mu_\eta^2+
    \mu_\sigma^2)+v^2
    (\lambda_{H\eta}^{\prime}+\lambda_{H\eta}
    +\lambda_{H\sigma}) +w^2 (\lambda_{\eta \phi}
    +\lambda_{\sigma \phi} )  \nonumber\\
    &&\pm S   \sqrt{\left[2 (\mu_\eta^2-
    \mu_\sigma^2)+v^2
    (\lambda_{H\eta}^{\prime}+\lambda_{H\eta}-
    \lambda_{H\sigma} )+w^2
    (\lambda_{\eta \phi} -\lambda_{\sigma\phi} )\right]^2+
    2 \kappa^2v^2 w^2}\Bigg\},
\end{eqnarray}
with
\begin{equation}
   S= \mathrm{sign}\left[2 (\mu_\eta^2-\mu_\sigma^2)+v^2
    (\lambda_{H\eta}^{\prime} +\lambda_{H\eta}-\lambda_{H\sigma})+w^2
    (\lambda_{\eta \phi} -\lambda_{\sigma \phi} )\right].
\end{equation}
\subsection{Gauge bosons}

From the covariant derivatives of the scalar fields 
\begin{eqnarray}
    D_\mu H &=& \left( \partial_\mu + i g \frac{\tau_i}{2} W_i^\mu + i  \frac{g^{\prime}}{2} B^\mu  \right)H ,\\
    D_\mu \phi &=& \left( \partial_\mu + i 3 g_L  B^{\prime \mu} \right)\phi \,. \nonumber
\end{eqnarray}
and the definitions
\begin{equation}
    W_\mu^{ \pm}=\frac{1}{\sqrt{2}}\left(W_{\mu 1} \mp i W_{\mu 2}\right) ,
    \end{equation}
In the leptophilic benchmark limit where gauge kinetic mixing between $B^{\mu}$ and $B^{\prime \mu}$ is neglected, we can readily identify the masses of the gauge bosons of the model from the kinetic term of the scalar fields after spontaneous symmetry breaking (SSB)
\begin{eqnarray}
    \left(D^\mu H\right)^\dagger \left(D_\mu H\right) + \left(D^\mu \phi\right)^\dagger \left(D_\mu \phi\right) &\supset \frac{1}{8} v^2 \left[\left( g W_3^\mu  - g^{\prime}B^\mu \right)^2+2 g^2
   W^{\mu +} W^{-}_\mu \right] \nonumber \\
   &  + \frac{1}{8} \left( 36  g_L^2 w^2 B^{\prime \mu } B^{\prime }_\mu \right) . 
\end{eqnarray}
The first term of the right-hand side can be written as 
\begin{equation}
    \frac{1}{8} v^2\left( g W_3^\mu - g^{\prime} B^\mu \right)^2 = \frac{ g^2 + g^{\prime 2}}{8} v^2 \left( \frac{g W_3^\mu - g^{\prime}B^\mu}{\sqrt{g^2 + g^{\prime 2}}} \right)^2=\frac{v^2}{8}\left(\begin{array}{ll}
    W_3^\mu & B^\mu
    \end{array}\right)\left(\begin{array}{cc}
    g^2 & -g g^{\prime} \\
    -g g^{\prime} & g^{\prime 2}
    \end{array}\right)\left(\begin{array}{c}
    W_{\mu 3} \\
    B_\mu
    \end{array}\right),
\end{equation}
which can be diagonalized by an orthogonal transformation of the form
\begin{equation}
    \left(\begin{array}{c}
W_{\mu 3} \\
B_\mu
\end{array}\right)=\left(\begin{array}{cc}
c_W & s_W \\
-s_W & c_W
\end{array}\right)\left(\begin{array}{c}
Z_\mu \\
A_\mu
\end{array}\right),
\end{equation}
where $c_W \equiv \cos{\theta_W}$, $s_W \equiv \sin{\theta_W}$, and $\theta_W$ is the Weinberg angle, given by
\begin{equation}
    \tan{2 \theta_W} = \frac{2 gg'}{g^2-g^{\prime 2}},
\end{equation}
and yielding
\begin{equation}
    \left(D^\mu H\right)^\dagger \left(D_\mu H\right) + \left(D^\mu \phi\right)^\dagger \left(D_\mu \phi\right)= \frac{g^2}{4}v^2 W^{\mu+} W_\mu^{-}  +\frac{\left(g^2+g^{\prime 2}\right)}{8}v^2 Z^\mu Z_\mu + \frac{9}{2} g_L^2 w^2 B^{\prime \mu } B^{\prime }_\mu,
\end{equation}
from where can read the mass terms 
\begin{equation}
\frac{g^2 v^2}{4} W^{\mu+} W_\mu^{-}+\frac{1}{2} \frac{\left(g^2+g^{\prime 2}\right) v^2}{4} Z^\mu Z_\mu + \frac{9}{2} g_L^2 w^2 B^{\prime \mu } B^{\prime }_\mu  =M_W^2 W^{\mu+} W_\mu^{-}+\frac{1}{2} M_Z^2 Z^\mu Z_\mu +\frac{1}{2} M_{Z^{\prime}}^2 Z^{\prime \mu } Z^{\prime }_\mu ,
\end{equation}
with $Z'^{\mu}=B'^{\mu}$ and 
\begin{equation}
    M_W^2 = \frac{g^2 v^2}{4}, \quad M_Z^2 = \frac{\left(g^2 + g^{\prime 2}\right)v^2}{4}, \quad M_{Z^{\prime}}^2 = 9 g_L^2 w^2.
\end{equation}
\paragraph{Kinetic mixing.}
For two Abelian factors the renormalizable gauge-kinetic Lagrangian allows a kinetic-mixing term,
\begin{equation}
\mathcal L_{\rm kin}\supset
-\frac14 B_{\mu\nu}B^{\mu\nu}
-\frac14 B'_{\mu\nu}B'^{\mu\nu}
-\frac{\epsilon}{2}\,B_{\mu\nu}B'^{\mu\nu}\,.
\end{equation}
Even if $\epsilon(\mu_0)=0$ at some reference scale $\mu_0$, it is in general regenerated by RGE running when
$\mathrm{Tr}(Y\,L)\neq 0$ \cite{Holdom:1985ag}. For the field content in Table~\ref{tab3211} this trace does not vanish
(e.g. the inert doublet $\eta$ carries both $Y$ and $L$), hence a small loop-induced $\epsilon$ is expected.
At leading order in small $\epsilon$, one can diagonalize the kinetic terms by a field redefinition
$B_\mu\to B_\mu+\epsilon B'_\mu$, which induces an additional coupling of $Z'_\mu\simeq B'_\mu$ to the hypercharge current,
i.e. $g_L L\to g_L L+\epsilon g' Y$. Therefore, for $\epsilon\neq 0$ the $Z'$ acquires suppressed couplings to quarks and
a small $Z$--$Z'$ mass mixing is induced after EWSB. In this work we adopt the simplifying benchmark $\epsilon(\mu_0)=0$ at
$\mu_0\sim w$ (leptophilic limit), corresponding to a UV completion with suppressed kinetic mixing; the loop-induced $\epsilon$
over the TeV-electroweak running interval remains small and does not qualitatively affect the DM and cLFV conclusions presented here.

 In the leptophilic limit ($\epsilon\to 0$) the $Z'$ does not couple to quarks, so hadron-collider production is strongly suppressed. For $\epsilon\neq 0$, loop-induced quark couplings arise via kinetic mixing and can be incorporated straightforwardly. Instead, the strongest constraints are imposed by LEP-II data \cite{Carena:2004xs} and imply 
\begin{equation}
w\geq 1.7\, \rm{TeV},
\end{equation}
roughly independent of the value of $g_L$ \cite{Schwaller:2013hqa}.

\section{Scotogenic Neutrino Masses}
\label{Sect:scotogenic}
Assuming that the parameter $\ell$ takes a value that leaves the new vectorlike generation of leptons unmixed with those of the SM, the following leptonic Yukawa interactions arise from the particle content and symmetries of the model:
\begin{equation}
        \begin{split}
        -L_Y=&y^e_{ij} \overline{L}_{iL} H e_{jR} + y^u_{ij} \overline{Q}_{iL} \widetilde{H} u_{jR} + y^d_{ij} \overline{Q}_{iL}H d_{jR} + y^\nu _{ij}\overline{iL}_L S_{jR} \widetilde{\eta} +
        h_{ia} \overline{S}_{iL} \nu_{a R} \sigma + M_S^D{}_{ij} \overline{S}_{iL} S_{jR}  \\
        &+y^{e'}_1\overline{e}''_{L}H^{\dagger}L''_{R}+y^{e'}_2\overline{L}'_{L}He'_{R}+y^{e'}_3\overline{e}''_{L}\phi e'_{R}+ 
        y^{e'}_4\overline{L}'_{L}\phi^{*} L''_{R} \\
        &+y^n_1\overline{n_L}\widetilde{H}^{\dagger}L''_{R}+ y^n_2\overline{L}'_{L}\widetilde{H}n_{R} +y^n_3\overline{n}_L\phi n_R +  \mathrm{h.c}.
        \end{split}
\end{equation}
with $\widetilde{H}\equiv i\tau_2H^{\star}$.  After SSB, SM quarks and SM charged leptons obtain their masses from the electroweak scale Higgs mechanism, while the new vectorlike lepton family develops masses at the $w$ scale.
 With the field assignments in Table \ref{tab3211}, the neutrino masses can be generated via the effective $dim=5$ operator similar to the Weinberg operator 
\begin{eqnarray}
    \mathcal{O}^D_5 &=& \frac{y_{ia}}{\Lambda_D} \phi^{\dagger} \overline{L}_{i\,L}\widetilde{H} \nu_{a\,R}+  \mathrm{h.c}.
\end{eqnarray}
leading after the symmetry breaking to two massive and one massless Dirac neutrinos. The smallness of the neutrino masses requires a large value of the new physics scale $\Lambda_D$. The tree-level UV completions of this operator, similarly to the Weinberg operator case, would require very heavy particles decoupled from the phenomenology.    In our model, this operator is realized at one-loop level with renormalizable interactions shown in Fig.~\ref{fig:diagrams}. The loop suppression allows for moderately large values for the mediators, which can yield nontrivial phenomenology, testable at colliders as well as in charged lepton flavor violation experiments like MEG II \cite{Meucci:2022qbh}, Mu2e \cite{Bernstein:2019fyh}, and COMET \cite{Moritsu:2022lem}, and provide both scalar and fermionic DM candidates in the scotogenic fashion.
To close the neutrino scotogenic loop in Fig.~\ref{fig:diagrams} we needed an extra set of fermion fields $S_{i\,L,R}$, which is vectorlike with respect to all the gauge group factors of the model. Their mass matrix $M^D_S=\mathrm{diag}(m_{S_1},m_{S_2},m_{S_3})$ is arbitrary but stable against radiative corrections due to chiral symmetry protection.   
The one-loop Dirac neutrino mass in Fig.~\ref{fig:diagrams} is finite and given by the expression 
\begin{figure}[ht]
    \centering
        \begin{tikzpicture}
            \begin{feynman}
            \vertex (i1);
            \vertex [right=2cm of i1] (a);
            \vertex [right=2cm of a] (b);
            \vertex [right=2cm of b] (c);
            \vertex [right=2cm of c] (f1);
            \vertex [below=0.001cm of b] (bb);
             \vertex [above=2cm of b] (tb1);
            \vertex [left=1.414cm of tb1] (l1);
             \vertex [above=1.414cm of l1] (l2){$\left\langle H \right\rangle$};
             \vertex [right=1.414cm of tb1] (r1);
               \vertex [above=1.414cm of r1] (r2){$\left\langle \phi\right\rangle$};
            \diagram* {
            i1 -- [fermion, edge label'=$L_L$] (a) -- [fermion, edge label'=$S_R$] (b) -- [fermion, edge label'=$S_L$] (c) -- [fermion, edge label'=$\nu_{aR}$] (f1),
            b -- [insertion=0.0] (bb),
            a -- [anti charged scalar, out=90, in=180, edge label=\(\eta\)] (tb1),
           tb1-- [anti charged scalar, out=0, in=90, edge label=\(\sigma\)] (c),
            tb1 -- [anti charged scalar, insertion=.9999] (l2),
            tb1 -- [anti charged scalar, insertion=.9999] (r2),
            };
            \end{feynman}
        \end{tikzpicture}
    \caption{One-loop diagram for neutrino masses.}\label{fig:diagrams}
\end{figure}
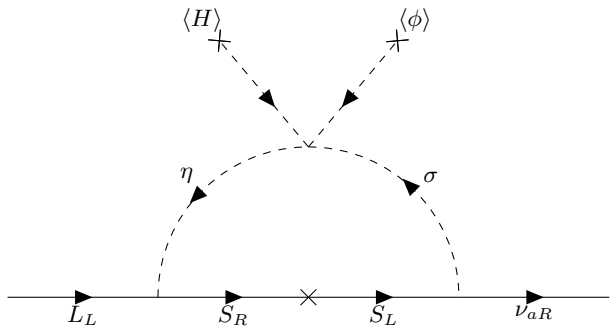\\
 {\color{blue} }
\begin{equation}
    \left(m_\nu\right)_{ij} = \frac{\sin{2\theta}}{32\pi^2}  \sum_{k} y_{ik}^\nu h_{kj} m_{S_k}\left[\frac{m_{\varphi^0_1}^2}{m_{\varphi^0_1}^2 - m_{S_k}^2} \ln \frac{m_{\varphi^0_1}^2}{m_{S_k}^2} - \frac{m_{\varphi^0_2}^2}{m_{\varphi^0_2}^2 - m_{S_k}^2} \ln \frac{m_{\varphi^0_2}^2}{m_{S_k}^2}\right].
    \label{Mnu}
\end{equation}
Here, $m_{S_k}$ represents the eigenvalues of the Dirac mass matrix $M_S^D$, while  $m_{\varphi^0_{1,2}}$ are the
 previously found eigenvalues of the mass eigenbasis of the rotated fields $\eta^0$ and $\sigma$, with $\theta$ 
 as their mixing angle. Notice that the resulting matrix is finite and has rank 2, since $h_{k3}=0$, predicting a massless state. Nevertheless, the resulting neutrino mass matrix has enough parametric freedom to successfully accommodate the neutrino oscillation experimental data.
 \begin{figure}[htbp]
\begin{center}
\includegraphics[width=0.95\linewidth]{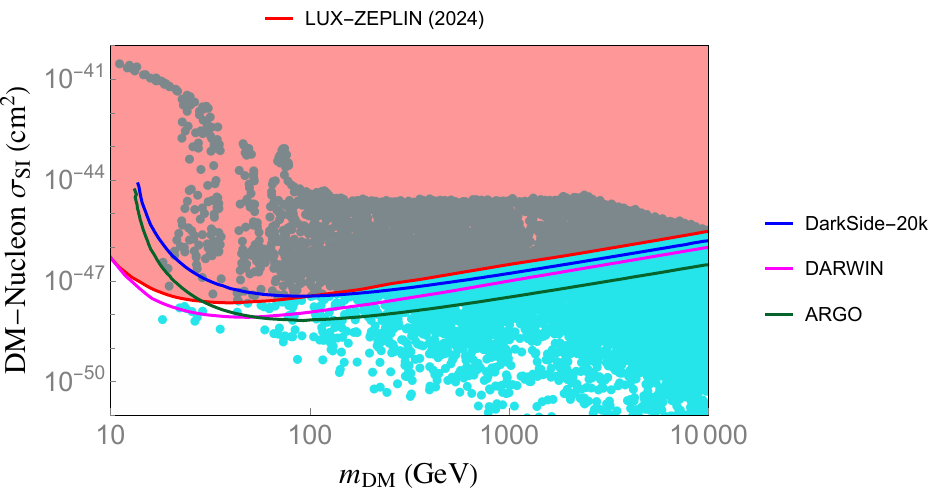}
\caption{Dark matter direct detection and relic abundance bounds. Each point represents a set of parameters that reproduces the correct relic abundance measured by PLANCK \cite{Planck:2018vyg}. Current limits from the LUX-ZEPLIN collaboration \cite{LZ:2024zvo}  are shown, together with future experiment projected sensitivities \cite{Billard:2021uyg,DarkSide-20k:2017zyg,Schumann:2015CPa}.}
\label{fig:DM}
\end{center}
\end{figure}

\section{Dark Matter}
\label{Sect:Dark}
The lightest electrically neutral particle transforming nontrivially under the remnant $\mathbb{Z}_6$ symmetry is automatically stable and can play the role of a WIMP DM candidate \cite{CentellesChulia:2019gic}. Here, we study the scenario in which DM is identified with the lightest neutral scalar of the dark sector. Assuming a small mixing angle $\theta$, the natural DM candidate is the complex scalar $\varphi^0_2$, as it is mostly composed by the electroweak singlet $\sigma$ and its coupling with the $Z$ boson is suppressed.

The restriction to small $|\theta|$ is phenomenologically motivated. Since $\eta$ carries electroweak charge, an appreciable $\eta^0$ component in the lightest neutral scalar would induce an unsuppressed coupling to the SM $Z$ boson, leading to a too-large elastic DM-nucleus scattering rate excluded by direct-detection searches. Therefore viable scalar WIMP DM points in this model are typically singlet-dominated, corresponding to $|\theta|\ll 1$ (cf. Eq.~(\ref{angle})). This limit is technically natural, since $\theta\to 0$ is achieved for $\kappa\to 0$ and/or in the hierarchical regime $w\gg v$.

In a generic Higgs portal DM framework, the parameter space for a single complex scalar is very narrow, essentially limited to near half of the Higgs mass, where resonant annihilation of dark matter into the Higgs boson takes place. Fortunately, in our scotogenic model, the parameter space for our WIMP DM candidate can be significantly widened by rescattering effects if the masses of the physical scalars are almost degenerate $m_{\varphi_1}\approx m_{\varphi_2}$ \cite{Kakizaki:2016dza}.

This enhancement is closely related to coannihilation/rescattering dynamics and is efficient provided the relative mass splitting satisfies $\Delta m/T_f\lesssim \mathcal O(1)$, with $T_f\simeq m_{\rm DM}/25$, i.e. $\Delta m/m_{\rm DM}\lesssim 0.1$--$0.2$. This motivates restricting our scan to a near-degenerate corridor.  For this analysis, we assume the hierarchy $w\gg v$, in which the masses of the new leptons required for gauge anomaly cancellation are large and decouple, and the mixing angle in Eq.(\ref{angle}) is small.
In Fig. \ref{fig:DM} , the results of our analysis in this scenario are presented. We have randomly varied the relevant Higgs portal couplings in the range $0<|\lambda_{H\sigma}|,|\lambda_{H\eta}|,|\lambda_{\eta\sigma}|<4\pi$ to ensure perturbativity, 
and we have scanned the mass parameters in the ranges $0<m_{\varphi_2}=m_{DM}<10^4\,\text{GeV}$, and $m_{\varphi_2}<m_{\varphi_1}<1.2m_{\varphi_2}$, with a mixing angle within $-0.01<\theta<0.01$. These scan ranges are therefore not fundamental assumptions of the model, but a targeted choice to efficiently explore the phenomenologically viable scalar-WIMP regime. Each point corresponds to a configuration that successfully reproduces the correct relic abundance $\Omega h=0.120\pm0.001$ \cite{Planck:2018vyg}.  As can be seen, there is a rather large parameter window below the current direct detection constraints imposed by the LUX-ZEPLIN experiment \cite{LZ:2024zvo}.

The model also admits a viable fermionic dark matter candidate, the Dirac fermion $S_1$. This field has a Higgs portal that contributes to the thermal relic density through the Yukawa couplings $y^\nu$ and $h$. This is an advantage over the conventional one-loop scotogenic Majorana neutrino mass models, where a single Yukawa coupling needs to take a relatively large value to produce the correct DM abundance, in tension with the experimental charged lepton flavor violation constraints \cite{Kubo:2006yx}, for models with extra annihilation channels to avoid cLFV in scotogenic models see~\cite{Bonilla:2019ipe,delaVega:2024tuu,CentellesChulia:2024iom}. In addition to the Higgs portal, $S_1$ can annihilate into SM fermion-antifermion pairs via the $t$ channel exchange of the $Z^{\prime}$ gauge boson, and in the secluded regime $m_{S_1}>m_{Z'}$ the annihilation may proceed via a pair of metastable on-shell $Z'$ particles, which ultimately decay to SM states \cite{Pospelov:2007mp}. The secluded regime is compatible with the LEP-II constraints for the mass of the $Z'$, discussed above, for $m_{S_1}$ values of order $\mathrm{TeV}$ \cite{delaVega:2022uko}.

\section{Charged lepton flavor violation}
\label{Sect:cLFV}
In what follows, we will analyze the phenomenological consequences of our model in  
charged lepton flavor violating (cLFV) decays. In particular, the process $\mu \rightarrow e\gamma $ arises at
one-loop level, thanks to the virtual exchange of the electrically charged
scalars $\eta ^{\pm }$, originating from the $\mathrm{SU(2)}_{W}$ inert doublet $\eta 
$ and the dark heavy neutral leptons $S_{i}$ ($i=1,2,3$).

The branching ratio for the $l_{\alpha }\rightarrow l_{\beta }\gamma $
decays is given by~\cite{Ma:2001mr, Toma:2013zsa, Vicente:2014wga,
Lindner:2016bgg} 
\begin{equation}
\text{BR}\left( l_{\alpha }\rightarrow l_{\beta }\gamma \right) =\frac{%
3\left( 4\pi \right) ^{3}\alpha _{\text{EM}}}{4\,G_{F}^{2}}\left\vert
A_{D}\right\vert ^{2}\,\text{BR}\left( l_{\alpha }\rightarrow l_{\beta }\nu
_{\alpha }\overline{\nu _{\beta }}\right) ,  \label{BRclfv}
\end{equation}%
where the form factor $A_{D}$ reads: 
\begin{equation}
A_{D}=\sum_{i=1}^{3}\frac{z_{\alpha i}z_{i\beta }^{\dagger }\,}{2(4\pi )^{2}}%
\frac{1}{m_{\eta ^{\pm }}^{2}}F\left( \frac{m_{S_{i}}^{2}}{m_{\eta ^{\pm
}}^{2}}\right) .
\end{equation}%
and the loop function $F\left( x\right) $ takes the form 
\begin{equation}
\label{eq:F-loop-1}
F\left( x\right) =\frac{1-6x+3x^{2}+2x^{3}-6x^{2}\log x}{6(1-x)^{4}}\,,
\end{equation}%
Here $z_{is}=\sum_{k=1}^{3}y_{ks}^{\nu }\left( V_{lL}^{\dagger }\right) _{ik}
$, where $V_{lL}$ is the rotation matrix that diagonalizes $%
M_{l}M_{l}^{\dagger }$ the charged lepton mass matrix. Besides that, $%
m_{\eta ^{\pm }}$ is the mass of the charged scalar component of the $\mathrm{SU(2)}_{W}$ inert doublet, while $m_{S_{i}}$ corresponds to the masses of the
dark heavy neutral leptons.

On the other hand, the branching ratio for 3-body decays $\ell_\alpha \to 3 \, \ell_\beta$ is given by~\cite{Toma:2013zsa, Vicente:2014wga}
\begin{align}
    \text{BR}\left(\ell_\alpha \to \ell_\beta \overline{\ell_\beta} \ell_\beta\right) &= \frac{3 (4\pi)^2 \alpha_\mathrm{EM}^2}{8\, G_F^2} \left[|A_{ND}|^2 + |A_D|^2 \left(\frac{16}{3} \log\left(\frac{m_\alpha}{m_\beta}\right) - \frac{22}{3}\right)\right. + \frac16 |B|^2 \nonumber\\
    & \quad \left. + \left(-2 A_{ND}\, A_D^* + \frac13 A_{ND} B^* - \frac23 A_D B^* + \mathrm{H.c.}\right)\right] \text{BR}\left(\ell_\alpha \to \ell_\beta \nu_\alpha \overline{\nu_\beta}\right). \label{eq:l3lBR}
\end{align}
The non-dipole photon penguin diagrams generate the form factor $A_{ND}$, which reads
\begin{equation} \label{eq:AND}
    A_{ND}=\sum_{k=1}^3\frac{z_{\beta k}^*\, z_{\alpha k}}{6 (4 \pi)^2} \frac{1}{m_{\eta^+}^2} G\left(\xi_k\right).
\end{equation}
Meanwhile, box diagrams yield the form factor $B$, expressed as
\begin{equation} \label{eq:B}
    e^2 B = \frac{1}{(4 \pi)^2 m_{\eta^\pm}^2} \sum_{k,\, l = 1}^3 \left[\frac12 D_1(\xi_k,\, \xi_l)\, z_{\beta l}^*\, z_{\beta l}\, z_{\beta k}^*\, z_{\alpha k} + \sqrt{\xi_k\, \xi_l} D_2(\xi_k,\, \xi_l)\, z_{\beta l}^*\, z_{\beta l}^*\, z_{\beta k}\, z_{\alpha k}\right],
\end{equation}
where the loop functions are defined by
\begin{align}
    G(x) &= \frac{2 - 9 x + 18 x^2 - 11 x^3 + 6 x^3 \log x}{6 (1-x)^4}\,, \\
    D_1(x,y) &= -\frac{1}{(1-x) (1-y)} - \frac{x^2 \log x}{(1-x)^2 (x-y)} - \frac{y^2 \log y}{(1-y)^2 (y-x)}\,, \\
    D_2(x,y) &= -\frac{1}{(1-x) (1-y)} - \frac{x \log x}{(1-x)^2 (x-y)} - \frac{y \log y}{(1-y)^2 (y-x)}\,.
\end{align}

\begin{figure}[tbp]
\centering
\includegraphics[width=9cm, height=8cm]{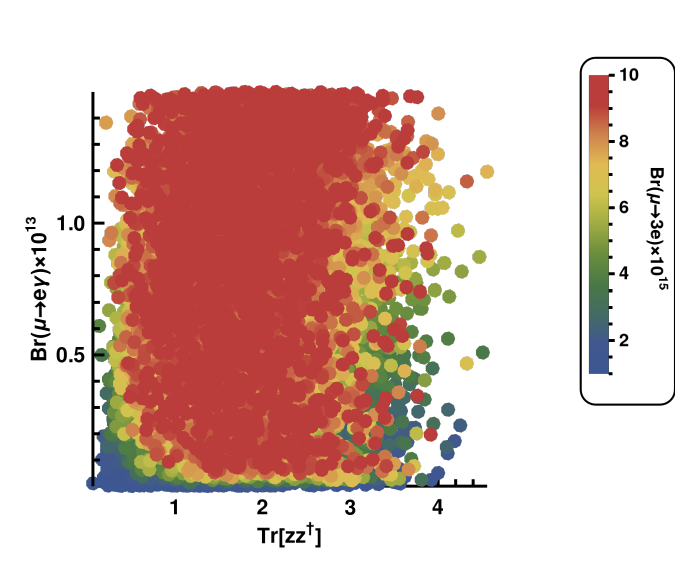}\includegraphics[width=9cm, height=8cm]{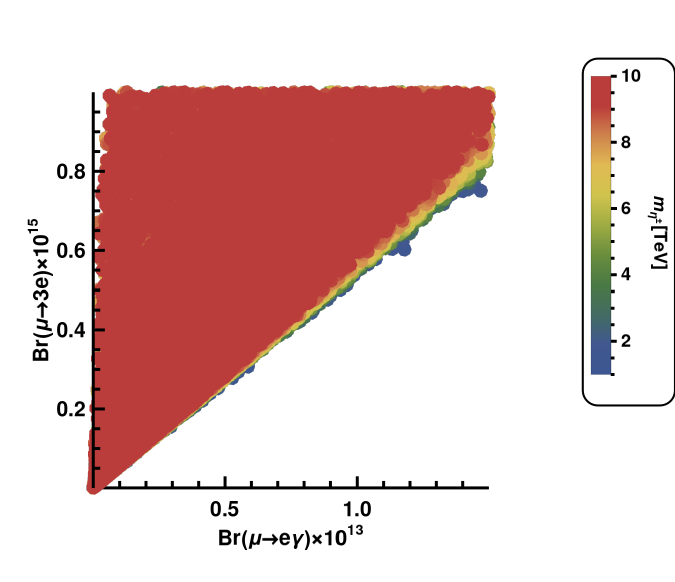}
\caption{Correlations between $Br\left(\protect\mu %
\rightarrow e\protect\gamma\right)$ and ${\rm Tr}\left(zz^{\dagger }\right)$ as well as between $Br\left(\protect\mu\rightarrow 3e\right)$ and $Br\left(\protect\mu %
\rightarrow e\protect\gamma\right)$. For the definitions see Sec.~\ref{Sect:cLFV}.}
\label{mutoegammas-1}  
\end{figure}

\begin{figure}[tbp]
\centering
\includegraphics[width=9cm, height=8cm]{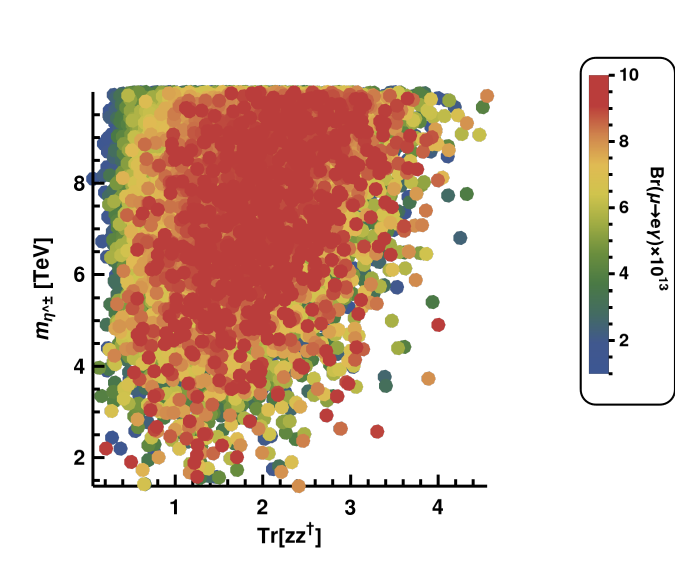}
\caption{Correlation between $m_{\eta^\pm}$ and $\mathrm{Tr}\left(zz^{\dagger }\right)$ for different values of the branching ratio of $\protect\mu \rightarrow e\protect\gamma$.}
\label{mutoegammas-2}
\end{figure}

The large numerical coefficient multiplying $\vert A _D \vert ^2$ in Eq.~\eqref{eq:l3lBR} typically renders the $A_D$ contribution more significant than $A_{ND}$ in $\mu \to eee$ decays. Furthermore, while $B$ depends on the fourth power of Yukawa couplings, $A _D$ and $A _{ND}$ scale with the second power. Consequently, for sufficiently small Yukawa couplings, $A _D$ dominates the amplitude, leading to a simple correlation between $\mu \to e e e$ and $\mu \to e \gamma$: the $\mu \to e e e$ rate becomes proportional to $\mu \to e \gamma$, albeit with a substantially smaller proportionality constant. This dipole-dominated scenario, commonly studied in the literature, implies that $\mu \to e e e$ is suppressed relative to $\mu \to e \gamma$, with the latter providing the most stringent constraints. However, for large Yukawa couplings, box contributions become important and cannot be ignored, particularly in the limits $m _{\eta ^+} \gg m_{S_i}$ or $m _{\eta ^+} \ll m_{S_i}$ ($i=1,2,3$), where $B$ is enhanced relative to $A _D$ due to the specific behavior of the $D _1$, $D_2$, and $F$ loop functions. In this regime, $\mu \to eee$ yields competitive constraints. Figure \ref{mutoegammas-1} displays the correlations between $Br\left(\protect\mu\rightarrow e\protect\gamma\right)$ decay and $\mathrm{Tr}\left(zz^{\dagger }\right)$ (left-panel) and between $Br\left(\protect\mu\rightarrow 3e\right)$ and $Br\left(\protect\mu\rightarrow e\protect\gamma\right)$ (right-panel) for different values of the electrically charged scalar masses and heavy Dirac neutrino masses. Figure \ref{mutoegammas-2} shows the correlation between $m_{\eta^\pm}$ and $\mathrm{Tr}\left(zz^{\dagger }\right)$ for different values of the branching ratio of $\protect\mu \rightarrow e\protect\gamma$. To generate the plots of Figs.~\ref{mutoegammas-1} and~\ref{mutoegammas-2}, we randomly varied the parameters in a range of values where the branching ratio for the $\mu\to e\gamma$ decay is below the experimental upper limit $1.5\times 10^{-13}$ \cite{MEGII:2025gzr}, whereas for the $\mu\to 3e$ decay we require that its corresponding branching ratio falls below the expected future experimental limit of $10^{-15}$. Let us note that the upper bound of $10^{-12}$ for the $\mu\to 3e$ branching ratio was established by the SINDRUM experiment in 1988~\cite{SINDRUM:1987nra}. However, it is worth mentioning that the first data-taking phase of the Mu3e experiment is scheduled for 2026. Its objective is to establish an upper limit on the $\mu\to 3e$ decay branching ratio of $10^{-15}$ with a projected sensitivity extending to the $10^{-16}$ \cite{COMET:2025sdw,Amarinei:2025ntv}. This is the reason why we impose in our numerical analysis that the branching ratio for the $\mu\to 3e$ decay to be below $10^{-15}$. Besides that, in generating the aforementioned plots, the entries of the neutrino Yukawa coupling matrix $z$ are taken in the range $[10^{-3},1]$, the electrically charged scalar masses are varied from $1$ to $10$ TeV, whereas the masses of the heavy Dirac neutrino seesaw messengers are taken as $0.5$ TeV $\leq  m_{S_1}\leq $ $2$ TeV, $m_{S_2}=m_{S_1}+\Delta$, $m_{S_3}=m_{S_1}+2\Delta$, with $\Delta=100$ GeV. These plots show that the vast majority of points in the parameter space compatible with charged-lepton-flavor-violating constraints correspond to charged-scalar masses near $10$ TeV. Moreover, there is a small number of parameter space points with charged scalar masses near $1$ TeV, for which the neutrino Yukawa couplings responsible for charged-lepton-flavor-violating decays are much smaller than unity. In summary, these plots show that our model satisfies the experimental constraints arising from charged-lepton-flavor-violation decays, and the obtained branching ratios for the $\mu\to e\gamma$ and $\mu\to 3e$ decays are within the current and future experimental sensitivity reach.       

\section{Summary and conclusions}
\label{Sect:Conclusions}

In this work, we have introduced the first scotogenic model based on a gauged lepton number $L$ symmetry, which is spontaneously broken by three units, yielding
a residual $\mathbb{Z}_6$, due to the existence of fractional charged fields. In this scenario, two of
the light active neutrinos acquire small masses through a
Dirac scotogenic mechanism at one-loop level, with radiative nature ensured by the preserved $\mathbb{Z}_6$ symmetry.  
The lightest electrically neutral state nontrivially charged under the remnant $\mathbb{Z}_6$ can be identified as a WIMP DM candidate, whose stability is protected by this $\mathbb{Z}_6$ symmetry. We have
studied the viability of the lightest scalar mediator in the scotogenic loop as the DM candidate, finding
that there is a sizable region of parameter space where the model can successfully reproduce the experimentally measured value of the dark matter relic abundance,  
while evading the current direct detection constraints. We have also analyzed the phenomenology of cLFV decays $\mu \rightarrow e\gamma $ and $\mu \rightarrow eee$, which arise at
one-loop level thanks to the presence of the charged 
scalars contained in the inert doublet needed for the scotogenic neutrino mass generation.  We have also found that our model is compatible with the constraints arising from charged lepton flavor-violating decays, with obtained rates within the experimental sensitivity reach.

\acknowledgments 
\noindent
A.E.C.H. is supported by ANID-Chile FONDECYT 1261103, 1241855, ANID – Millennium Science Initiative Program $ICN2019\_044$,  ANID CCTVal CIA250027 and ICTP through the Associates Programme (2026-2031). S.K. is supported by by ANID-Chile FONDECYT 1230160, Milenio-ANID-ICN2019$\_$044. E.P. is supported by DGAPA UNAM Grant No. PAPIIT-IN111625 and Fundaci\'on Marcos Moshinsky. The relic abundance and direct detection constraints were calculated using the micrOMEGAS package \cite{Alguero:2023zol} at GuaCAL (Guanajuato Computational Astroparticle
Lab).

\appendix

\section{ANOMALY CANCELLATION}
\label{App:anomaly}

In this Appendix we explicitly show how the potential anomalies from gauged lepton number vanish with the inclusion of the extra fields
\begin{itemize}
\item $[SU(2)_W]^2U(1)_L$
\begin{equation}
\begin{split}
\sum_{\text{ferm doub}}L_L-\sum_{\text{ferm doub}}L_R&=3\times 1\times 1+1\times1\times\left(\ell-3\right)-1\times1\times\ell=0.
\end{split}
\end{equation}
\item $[U(1)_L]^3$
\begin{equation}
\begin{split}
\sum_{\text{fermions}}L^3_L-\sum_{\text{fermions}}L^3_R=&3\times 1\times 2\times 1^3+1\times 1\times 2\times\left(\ell-3\right)^3+1\times 1\times 1\times\ell^3\\&+1\times 1\times 1\times\ell^3+1\times 1\times 1\times\left(\frac{1}{2}\right)^3\\&-3\times 1\times 1\times 1^3-2\times 1\times 1\times 4^3-1\times1\times1\times(-5)^3-1\times1\times1\times(\ell-3)^3\\&-1\times1\times1\times(\ell-3)^3-1\times2\times1\times\ell^3-1\times 1\times 1\times\left(\frac{1}{2}\right)^3=0.
\end{split}
\end{equation}
\item $[\text{Grav}]^2U(1)_L$
\begin{equation}
\begin{split}
\sum_{\text{fermions}}L_L-\sum_{\text{fermions}}L_R=&3\times 1\times 2\times 1+1\times 1\times 2\times\left(\ell-3\right)+1\times 1\times 1\times\ell\\&+1\times 1\times 1\times\ell+1\times 1\times 1\times\left(\frac{1}{2}\right)\\&-3\times 1\times 1\times 1-2\times 1\times 1\times 4-1\times1\times1\times(-5)-1\times1\times1\times(\ell-3)\\&-1\times1\times1\times(\ell-3)-1\times2\times1\times\ell-1\times 1\times 1\times\left(\frac{1}{2}\right)=0.
\end{split}
\end{equation}
 \item $[U(1)_Y]^2U(1)_L$   
\begin{equation}
\begin{split}
\sum_{\text{fermions}}Y^2_LL_L-\sum_{\text{fermions}}Y^2_RL_R=&   3 \times 1 \times 2 \times \left(-\frac{1}{2}\right)^2 \times 1   + 
  1\times 1\times 2 \times\left(-\frac{1}{2}\right) ^2\times (\ell-3)  \\&+ 1\times 1 \times 1\times(-1)^2 \times \ell - 
  3 \times 1 \times 1 \times(-1)^2 \times 1 \\&-  1\times 1\times2\times \left(-\frac{1}{2}\right)^2 \times\ell- 1\times 1 \times 1\times(-1)^2 \times(\ell-3)  =0.
\end{split}
\end{equation}
\item $[U(1)_Y]U(1)_L^2$ 
\begin{equation}
\begin{split}
\sum_{\text{fermions}}Y_LL^2_L-\sum_{\text{fermions}}Y_RL^2_R=&  3 \times 1 \times 2 \times \left(-\frac{1}{2}\right) \times 1^2   + 
  1\times 1\times 2 \times\left(-\frac{1}{2}\right) \times (\ell-3)^2  \\&+ 1\times 1 \times 1\times(-1) \times \ell^2 - 
  3 \times 1 \times 1 \times(-1) \times 1^2 \\&-  1\times 1\times2\times \left(-\frac{1}{2}\right) \times\ell^2- 1\times 1 \times 1\times(-1) \times(\ell-3)^2 =0.
\end{split}
\end{equation}
\end{itemize}

\bibliographystyle{utphys}
\bibliography{bibliography}
\end{document}